# Nanoscopic distribution of VAChT and VGLUT3 in striatal cholinergic varicosities suggests colocalization and segregation of the two transporters in synaptic vesicles

Paola Cristofari [1]†, Mazarine Desplanque [1]†, Odile Poirel [1], Alison Hébert [1], Sylvie Dumas [2], Etienne Herzog [3], Lydia Danglot [4,5,6], David Geny [5], Jean-François Gilles [7], Audrey Geeverding[8], Susanne Bolte [7], Alexis Canette [8], Michaël Trichet [7], Véronique Fabre [1], Stéphanie Daumas [1], Nicolas Pietrancosta [1,9], Salah El Mestikawy [1,10] and Véronique Bernard [1,11*]

†Both authors contributed equally to this work

[1] Sorbonne Université - CNRS UMR 8246 - INSERM U1130 - Neuroscience Paris Seine - Institut de Biologie Paris Seine (NPS - IBPS), F-75005 Paris, France.

[2] Oramacel, F-75006 Paris, France.

[3] Univ. Bordeaux, CNRS, Interdisciplinary Institute for Neuroscience, IINS, UMR 5297, F-33000 Bordeaux, France.

[4] Université de Paris, Institute of Psychiatry and Neuroscience of Paris (IPNP), INSERM U1266, Membrane Traffic in Healthy & Diseased Brain, F-75014 Paris, France.

[5] Université de Paris, Institute of Psychiatry and Neuroscience of Paris (IPNP), INSERM U1266, NeurImag Imaging facility, F-75014 Paris, France.

[6] GHU PARIS Psychiatrie & Neurosciences, F-75014 Paris, France.

[7] Imaging facility of the Institut de Biologie Paris- Seine (IBPS) - Sorbonne Université, F-75005 Paris, France.

[8] Sorbonne Université, CNRS, Institut de Biologie Paris-Seine (IBPS), Service de microscopie électronique (IBPS-SME), F-75005, Paris.

[9] Sorbonne Université - CNRS UMR 7203 – Laboratoire des BioMolécules, F-75005 Paris, France.

[10] Douglas Mental Health University Institute, Department of Psychiatry, McGill University, Montreal, Canada.

[11] Lead contact

**Correspondence**: Véronique Bernard (veronique.bernard@inserm.fr)

**Abstract**

Striatal cholinergic interneurons (CINs) use acetylcholine (ACh) and glutamate (Glut) to regulate the striatal network since they express vesicular transporters for ACh (VAChT) and Glut (VGLUT3). However, whether ACh and Glut are released simultaneously and/or independently from cholinergic varicosities is an open question. The answer to that question requires the multichannel detection of vesicular transporters at the level of single synaptic vesicle (SV). Here, we used super-resolution STimulated Emission Depletion microscopy (STED) to characterize and quantify the distribution of VAChT and VGLUT3 in CINs SVs. Nearest-neighbor distances analysis between VAChT and VGLUT3-immunofluorescent spots revealed that 34 % of CINs SVs contain both VAChT and VGLUT3. In addition, 40 % of SVs expressed only VAChT while 26 % of SVs contain only VGLUT3.

These results suggest that SVs from CINs have the potential to store simultaneously or independently ACh and/or Glut. Overall, these morphological findings support the notion that CINs varicosities can signal with either ACh or Glut or both with an unexpected level of complexity.

**Introduction**

The striatum regulates reward- and habit-guided behaviors as well as locomotor activity (Graybiel, 2008; Palmiter, 2008) and is involved in a wide range of neurological and psychiatric disorders (Crittenden et al., 2014; Florio et al., 2018). Striatal GABAergic medium spiny output neurons are modulated by dopaminergic inputs as well by striatal cholinergic interneurons (CINs) (Witten et al., 2010; Lim et al., 2014). CINs represent ≈1% of striatal neurons but form an extensively ramified network (Morris et al., 2004; Cragg, 2006; Goldberg and Reynolds, 2011). CINs express vesicular transporters for acetylcholine (VAChT) and for glutamate (VGLUT3) and consequently regulate the striatal network with both acetylcholine (ACh) and glutamate (Glut) (Gras et al., 2002, 2008; Higley et al., 2011). In the striatum, VGLUT3 enhances ACh accumulation and cholinergic transmission (Gras et al., 2008). Similarly, recently, it was found that $Zn^{2+}$ facilitate Glut vesicular accumulation through the presence of two transporters ZnT3 and VGLUT1 on the same vesicles (Upmanyu et al., 2022). This presynaptic mechanism named "vesicular synergy" has also been reported in serotonergic, dopaminergic and GABAergic terminals (Hnasko et al., 2010; El Mestikawy et al., 2011; Amilhon et al., 2010; Frahm et al., 2015; Trudeau and El Mestikawy, 2018; Zander et al., 2010). The current mechanistic explanation of vesicular synergy is based on a VGLUT-dependent acidification of SVs due to the intralumenal accumulation of Glut (Gras et al., 2008; Amilhon et al., 2010; Hnasko et al., 2010; El Mestikawy et al., 2011; Frahm et al., 2015). This model and the fact that immuno-isolated VGLUT3-positive SVs from rat striatum adsorbed VGLUT3- and VAChT-positive vesicles suggest (but does not prove) that VGLUT3 and VAChT are expressed on the same population of SVs. However, several recent findings and indirect evidence conflict with this hypothesis. For example, cholinergic neurons from the medial habenula, projecting to the interpeduncular nucleus, corelease ACh and Glut (Ren et al., 2011; Frahm et al., 2015; Souter et al., 2022). Optogenetic single pulse stimulation of interpeduncular cholinergic terminals triggers glutamatergic currents whereas their tetanic stimulation elicit cholinergic currents (Ren et al., 2011). Therefore, these cholinergic terminals have the capacity to independently release ACh or Glut depending on their firing frequency. This finding suggests that in the IPN, ACh and Glut are stored, at least partially, in independent pools of SVs. In addition, a recent study demonstrated, using single-vesicle imaging techniques, that only a minority of SVs from total brain extract contains two transporters (Upmanyu et al., 2022). Therefore, whether a single SV from CINs can express VAChT or / and VGLUT3 and consequently store independently or co-store ACh or Glut remains to be established.

One way to test our hypotheses is to visualize vesicular transporters to determine their distribution at the level of single SVs. Here, we used nanoscale imaging with STimulated Emission Depletion microscopy (STED) to assess the relative distribution of VAChT and VGLUT3 in CINs SVs. We show that striatal cholinergic varicosities contain three subpopulations of SVs expressing either VAChT or VGLUT3, or both. These findings provide morphological evidence that open the way to a deeper understanding of mechanisms underlying ACh/Glut co-transmission by CINs.

## Materials and Methods

### Animals

Animal care and experiments were conducted in accordance with the European Communities Council Directive for the Care and the Use of Laboratory Animals (Regulation EU 2019/1010) and in compliance with the *Ministère de l'Agriculture et de la Forêt, Service Vétérinaire de la Santé et de la Protection Animale* (authorization number 01482.01). All efforts were made to minimize the number of animals used in the course of the study and to ensure their well-being. The animals were housed in a temperature-controlled room (21 +/- 2°C) with *ad libitum* access to water and food under a 12 h light/dark cycle (lights on 7:30 A.M. to 7:30 P.M.).

### Experimental design and statistical analysis

Experiments were performed in two groups of animals aged of 2 to 3 months. The first group was used for the analysis of the localization of VAChT and VGLUT3 in CINs cell bodies and varicosities by immunohistochemistry on striatal sections. The second group was used for the visualization of SVs by scanning electron microscopy (SEM) and for the localization of VAChT and VGLUT3 by immunohistochemistry on striatal isolated SVs. This latter group was subdivided into two subgroups: wild type mice (genetic background C57BL/6) and VGLUT3 knock out mice (VGLUT3 $^{-/-}$) (Gras et al., 2008).

Statistical analyses were performed using Graphpad Prism 9 (San Diego, CA 92108, USA). Data were tested for normality using D'Agostino-Pearson omnibus test. Comparisons were made by appropriate non-parametric tests. The Wilcoxon matched-pairs signed rank test compared the mean diameter of immunofluorescent spots for VAChT and VGLUT3 from confocal and STED microscopies. The same test was used to compare the mean diameter of immunofluorescent spots for VAChT and VGLUT3 with and without deconvolution. The Kruskal-Wallis test followed by the Dunn's post-hoc test compared the number of VAChT,

VGLUT3 or VAChT/VGLUT3 immunopositive spots per surface of microscope slide in isolated vesicles. The Kolmogorov-Smirnov test was used to compare the frequency distribution of the nearest neighbour distances (NNDs) for VAChT-VGLUT3 and VGLUT1-VGLUT3 or VGLUT2-VGLUT3-immunopositive spots. The significance level was set at 5%. The size and type of individual samples, n, for given experiments is indicated and specified in the results and extending data sections, in figure legends and in tables. Data were expressed as means $\pm$ SEM and $p$ values $< 0.05$ were considered as statistically significant. The experiments were repeated at least 3 times. Asterisks indicate p values as follows: *$p < 0.01$, **$p<0.0001$.

**Tissue preparation**

Brain sections: For immunohistochemical experiments, mice were sedated with sodium pentobarbital (intraperitoneal route, 10 mg / 20 g, Ceva Santé Animale, France), then perfused intracardiacally with 2% paraformaldehyde (PFA) in 0.1 M PBS, pH = 7.4 at 4 °C for 15 min. The brains were removed and post-fixed overnight in 2 % PFA, then transferred to PBS containing 0.03 % sodium azid (Sigma-Aldrich, Saint-Louis, Missouri, USA). Fifty-µm thick brain sections including the striatum were made with a vibratome (VT1000S, Leica Biosystems, Wetzlar, Germany). The sections were then stored in PBS / sodium azid at 4 °C until use. For dFISH experiments, mice were euthanized by cervical dislocation. Brains were removed, rapidly frozen in cold isopentane (-30 °C / -35 °C) and sectioned at the level of striatum on cryostat at 16 µm thickness.

Synaptic vesicles: In brain sections, SVs are tightly packed in CINs varicosities. This may impede of the efficient penetration of reagents including antibodies and thus lead to an underestimation of the quantity of detected transporters. To overcome this concern, isolated striatal mouse SVs were used for the visualization of SVs by SEM and for the analysis of the nanoscopic localization of VAChT, VGLUT3, VGLUT1, VGLUT2 and by immunohistochemistry. SVs were prepared as previously described (adapted from Huttner et al. 1983). In brief, mice were sacrificed by cervical dislocation. Three SVs preparations with striata from WT mice and one with VGLUT3$^{-/-}$ mice were performed. For each preparation, striata from 30 WT mice or 30 VGLUT3$^{-/-}$ mice were dissected and homogenized in ice-cold sucrose buffer (0.32 M; 4 mM KCl, 4 mM MgSO4 and 10 mM HEPES–KOH, pH 7.4). Striatal SVs were purified by differential centrifugations. Supernatants from two successive centrifugations at 1.000g (10 min) were collected and centrifugated again twice at 12.500g (15 min). The resulting pellets containing synaptosomes were homogenized in sucrose buffer. Synaptosomes were then lysed by osmotic shock in iced water, homogenized and centrifugated at 25.000g (20 min). The supernatant was collected and centrifugated at 165.000g (2 hrs). The resulting pellet containing the SVs was homogenized in sucrose buffer and stored at -80 °C. Protein levels were quantified by the Bradford essay. Before use, SVs were unfrozen, washed and centrifugated 3 times in PBS. The SV solution (25 µg of proteins in 300 µl of PBS) was dropped off on glass slides (Ibidi 12 well chamber slide; Biovalley, Nanterre, France) on the surface of which the SVs settled for 30 min at 37 °C and processed either for SEM observations or for immunohistochemistry and observation under the confocal and STED microscopes.

**Immunohistochemistry**

*Antibodies:* VAChT was detected with an anti-VAChT polyclonal antiserum raised in guinea pig (Gras et al., 2008). VGLUT3 was detected with an anti-VGLUT3 polyclonal antiserum raised in rabbit. The specificity of VAChT and VGLUT3 labeling was assessed by the disappearance of the staining in VAChT and VGLUT3 null mice (data not shown; Gras et al., 2002, 2008; Amilhon et al., 2010; Guzman et al., 2011). VGLUT1 or VGLUT2 were detected with polyclonal antisera raised in goat or guinea-pig, respectively. See Table 1 for details. All antibodies against vesicular transporters recognize epitopes localized at the cytoplasmic side of SV's membrane.

*Co-detection of VAChT and VGLUT3 on brain sections:* Free-floating sections were first permeabilized with 0.25 % TritonTM-X100 (TX, Sigma-Aldrich) and incubated in blocking solution (1X PBS containing 0.25 % TX, 4 % normal goat serum (NGS, Thermo Fisher Scientific, Waltham, Massachusetts, USA)) for 30 min at room temperature, to permeabilize membranes and saturate non-specific sites. Co-detection of two proteins was carried out by incubation in a mixture of primary antibodies recognizing each of the proteins of interest, diluted in a solution of 1X PBS-0.25 % TX - 1 % NGS. After incubation overnight at room temperature with stirring, sections were washed in 1X PBS-0.25 % TX, then incubated for 2 hrs with the secondary antibodies coupled to different fluorophores (Alexa-594 and Abberior-Star 635P; dilution: 1: 100 in 1X PBS-0.25% TX). All fluorophores used are suitable for confocal and STED microscopy. They were washed again in 1X PBS-0.25 % TX, rinsed in 1X PBS, then mounted on slides with ProLong Gold or ProLong Diamond Antifade Reagent (Thermo Fisher Scientific). Sections were then observed and acquired in a STED super resolution microscopy (LEICA SP8 - STED 3X microscope equipped with a 775nm depletion laser, respectively (Leica Microsystems, Wetzlar, Germany).

*Co-detection of VAChT and VGLUT3, or VGLUT3 and VGLUT1 or VGLUT2 in isolated SVs:* Immunohistochemistry on isolated vesicles was performed on SVs adsorbed on glass slides (ibidi 12 well chamber slide; Biovalley). SVs were incubated in blocking solution (1X PBS containing 4% NGS (or normal donkey serum (NDS) when the primary antibody was made in goat) for 15 min at 37°C to saturate non-specific sites. Co-detections were carried out by incubation in a mixture of primary antibodies recognizing each of the proteins of interest, diluted in a solution of 1X PBS-1% NGS (or NDS) for 2 hrs at 37°C. The SVs were washed in 1X PBS then incubated for 1 hr with the secondary antibodies coupled to Alexa-594 and Abberior-Star 635P fluorophores (Abberior GmbH, Göttingen Germany; dilution: 1: 1000 in 1X PBS). SVs were finally washed in 1X PBS, fixed in 2 % PFA for 5min, washed and then mounted on slides with ProLong Diamond Antifade Mountant (P36961, Thermo Fisher Scientific). Stainings were observed and acquired in a STED super resolution microscope (LEICA SP8 - STED 3X, equipped with a 775nm depletion laser; Leica Microsystems).

**Scanning electron microscopy (SEM)**

SEM was used to analyze the morphology of SVs and to check the absence of clusterization. After wash in PBS (see above), the vesicle solution (5-50 μg / 300 μl PBS) was dropped off on glass slides (Ibidi 12 well chamber slide; Biovalley, France) where the vesicles settled at their surface for 30 min at 37 °C. Then, SVs were fixed in 2 % glutaraldehyde in cacodylate buffer 0.1 M. After washes in cacodylate buffer 0.1 M, SVs were post-fixed with 1 % osmium tetroxide in 0.1 M cacodylate buffer for 10 min, washed and dehydrated in ascending concentrations of ethanol (50 % to 100 %), then in hexamethyldisilazane, before drying by evaporation under primary vacuum. SVs were covered by a thin layer (5 nm) of platinum using a Sputter Coater Leica EM ACE600 (Leica Microsystems, Wetzlar, Germany), in order to enhance both the contrast and the electrical conductivity of the samples. They were observed under a Field Emission scanning electron microscope GeminiSEM 500 (Zeiss microscopy, Jena, Germany) operating in high vacuum, at 3 kV, with a 20 μm aperture, high current mode, and a working distance around 2.5 mm. Secondary electrons were collected with the "inLens" detector. Images were acquired with a 1024 x 768 definition, with a pixel dwell time of 1.6 μs and a line averaging of 10. The diameter of SVs was measured using the Fiji software.

**Confocal and STED microscopy observations**

Confocal and STED microscopy were used to codetect VAChT and VGLUT3 in striatal sections and in isolated SVs. Imaging was performed at the surface of brain sections (first 10µM) where the tissue permeabilization is the most efficient. Confocal microscopy imaging was performed using a Leica TCS SP5 (Leica Microsystems) at the Imaging facility of the IBPS. For that, images were acquired in the dorsolateral striatum (DLS) using a hybrid detector (HyD) and a x63 oil immersion objective, NA 1.4, with a zoom x5. STED imaging was carried out with a Leica SP8 - STED 3X microscope (Leica Microsystems) within NeurImag facility (IPNP). Sections were observed with a x93/1.3 NA glycerol immersion objective with two hybrid detectors (HyDs) within DLS. AlexaFluor 594 and AbberiorStar 635P-coupled secondary antibodies were excited at 590 nm and 631 nm respectively with a pulsed white laser, and fluorescence was depleted using a pulsed depletion laser (775nm). Typically, images of 1024 x 1024 pixels were acquired with a x5 magnification resulting in a pixel size in the range of 20-25 nm for tissue and 12-15 nm for isolated SVs. Using the STED microscope, the same fields may be observed alternatively with the confocal and with the STED modes. STED images were deconvolved with the Huygens 3.7 software (Scientific Volume Imaging, Hilversum, Netherlands), which permits the recovery of objects that are degraded by optical blurring and noise. Deconvolution allows also to improve the signal to noise ratio and the resolution. We used an automatically computing theoretical PSF based on the microscopic parameters including the Microscope Type, the numerical aperture, the lens and medium refractive indexes, the channel label, the excitation and emission wavelengths. We used the Classical Maximum Likelihood Estimation (CMLE) algorithm. Signal-to-noise ratio also called the R-parameter was set to a value of 2.6. The number of iterations was 25. The background intensity was averaged from the pixels with lowest intensity. We checked that deconvolution did not create artifacts and we measured the diameter of the spots obtained under the STED microscope before and after deconvolution (Supplementary Figure 2). For the purpose of minimizing noise, we applied a Gaussian 2D kernel with a Gaussian width, σ, to the raw and deconvolved images. In our case, we have used the Gaussian Blur filtering function in the Fiji compilation of ImageJ 1.47N with a Gaussian width of 1. The full width at half maximum corresponding to the diameter of the fluorescent spots was calculated with the Fiji plugin "plot profile". Finally, images were analyzed using ImageJ and mounted with Adobe Photoshop, Adobe Indisign and Abobe illustrator.

Therefore, as expected, deconvolution increased the resolution of STED images without introducing major alteration of the fluorescent signal as previously shown (Wegel et al., 2016).

**Quantification**

*Quantification of the diameter of SVs from scanning electron microscopy pictures:* The quantification of the SVs diameter was performed using the Fiji software. Since the SVs were covered by a 5 nm layer of platinium, we substracted 10 nm to the measured diameter (5 nm on each side of SVs).

*Quantification of the proportion of the spots displaying couples of immunolabelings separately or colabelings on STED images on striatal sections:* The STED microscopy images on brain sections were manually analyzed to quantify the number of immunofluorescent spots for VAChT or VGLUT3. Labelled cholinergic varicosities were easily identified as clusters of VAChT and VGLUT3-fluorescent spots (see Fig. 1D,D',C'). The surface of the cluster was measured using the Fiji software. Three classes of spots were visually identified and counted: one for each VAChT or VGLUT3 labeling, separately, and one displaying both VAChT and VGLUT3 labeling. The density of VAChT, VGLUT3 and VAChT/VGLUT3 spots was calculated (number of spots per varicosity/surface of varicosity).

Clustered fluorescent spots were considered as specific spots in varicosities whereas isolated fluorescent spots were considered as background and were excluded from the quantification.

*Batch analysis quantification of NNDs between immunofluorescent spots for VAChT and VGLUT3 in striatal sections observed under a STED microscope*: NNDs between VAChT and VGLUT3 spots in STED pictures were estimated using a dedicated automatized program using Icy software and "protocol" within NeurImag. Briefly, VAChT and VGLUT3 clusters were segmented through wavelet analysis using spot detector plugin as previously described (Lagache et al., 2018). Spots with a surface area bigger than 2 pixels were considered in further analysis. Mass center of spots were retrieved and thanks to X Y coordinates, Euclidian distance between each couple of VAChT-VGLUT3 clusters were measured. For each VGLUT3 clusters, NND to other VAChT spots was extracted spots by spots and retrieved for each picture in separated excel file rows. Matrix of NND was then filtered in excel to only retrieve VAChT-VGLUT3 pairs contained in a radius of 1000 nm. Frequency distribution of filtered NND was then plotted in Prism 9 software with a bin of 10 nm.

*Quantification of the NNDs between immunofluorescent spots for VGLUT3 and VAChT, VGLUT3 or VAChT and VGLUT1 or VGLUT2, on STED images with isolated vesicles*: Preparations of isolated SVs were observed under the STED microscope. Images were deconvolved and the fluorescent spots diameters were analyzed using the plot profile plugin of the Fiji software. Images were analyzed using the DiAna plugin of the ImageJ software (Gilles et al., 2017). Briefly, the plugin quantified the number of VAChT and VGLUT3 immunofluorescent spots and calculate their spatial coordinates. The plugin was also used to perform automatized analysis of the NNDs between fluorescent spots for VGLUT3 and their closest fluorescent spot for VAChT, VGLUT1 or VGLUT2 on purified SVs preparations. The analysis is based on the automated object-based distance analysis. The center-to-center

distance, i.e. the distance between the center of the two immunofluorescent spots, also called NND is equal to 0 if the two objects colocalize. Spots with a surface area bigger than 10 pixels were considered in further analysis.

Cristofari et al. VAChT and VGLUT3 distribution in CINs

**Results**

**VAChT and VGLUT3 distribution in CINs**

VAChT and VGLUT3 were first visualized with classical confocal microscopy in cell bodies and axonal varicosities of CINs (Figure 1A,B). Low magnification images showed that VAChT and VGLUT3 were codetected in the same perikaryia (Figure 1A) and varicosities (Figure 1B). Surprisingly, at higher confocal magnification, we noticed that VAChT and VGLUT3 immunolabeling mainly did not overlap in axonal varicosities (Figure 1C). To confirm this observation, the same varicosity was visualized with STED super-resolution microscopy (Figure 1C'). STED imaging bore out that VAChT-positive and VGLUT3-positive fluorescent puncta were largely non-overlapping (Figure 1C',D,D'). Manual quantification confirmed that VAChT and VGLUT3 fluorescent spots were mostly dissociated. Only a very small proportion strictly overlapped and probably represent VAChT and VGLUT3 fluorescent spots whose centrers are located ≈ 0-20nm away from each other (1%, Figure 1E,F). This finding suggests that VAChT and VGLUT3 are frequently distant from each other.

In order to analyze more precisely the distances between VAChT and VGLUT3 in CINs varicosities from the dorsolateral striatum, we performed an automatic batch analysis of the NNDs between the centers of VAChT- and VGLUT3-immonopositive spots (Figure 1G). The frequency histogram revealed a peak with NNDs between VAChT and VGLUT3 centered around $40.0 \pm 5$ nm (Figure 1G). The size of a SVs is estimated to be ≈50 nm and the size of the complex composed of primary/secondary antibodies and a fluorochrome to ≈45nm (Figure 1H; Früh et al., 2021). A cut-off of 95 nm was used to estimate the proportion of cholinergic VAChT-immunolabeled SVs containing VGLUT3 (< 95 nm) or not (> 95 nm) (Figure 1H). In striatal sections, 30% of VAChT-immunopositive spots were located between 0 and 95 nm from their closest VGLUT3 spot (Figure 1I). Similar results were found in the dorsolateral striatum, dorsomedial striatum and nucleus accumbens (data not shown). Our finding suggests that about one third of VAChT and VGLUT3 molecules could be located on the same SVs.

**VAChT and VGLUT3 distribution in purified striatal SVs**

In striatal sections, SVs are very tightly packed. Different parameters including the lack of penetration of reagents in tissue and the steric hindrance of the primary/secondary antibody/fluorochrome complex (Figure 1H) may lead to misinterpretation of the precise localization of fluorescent spots. To overcome this difficulty and to better estimate the average NND between VAChT and VGLUT3 in less intricated conditions than in a varicosity, we then used purified SVs preparations from the mouse striatum. Previous studies suggested that purified SVs might be associated in clusters (Huttner et al., 1983). This clusterization could bias co-localization analyses in purified striatal SVs preparation. Thus, we first processed preparation of striatal SVs for Scanning Electron Microscopy (SEM) to analyze their dispersion. Under a SEM microscope, our preparations contained SVs that were not clustered (Figure 2A). The mean diameter of SVs was $49.7 \pm 0.8$ nm (Quantification from 484 SVs; Figure 2B). This result shows that our striatal SV preparations are adapted for colocalization analysis.

Then, we used immunofluorescent detection of VAChT and VGLUT3 with a confocal or STED microscope to assess their distribution in purified striatal SVs (Figure 2C,C',D,E-G). As expected, relatively to confocal microscopy, STED microscopy increased the resolution of VGLUT3 immunofluorescent spots (Figure 2C,C'). Indeed, the diameter of VGLUT3-immunofluorescent puncta was $260 \pm 3$ nm at the confocal level compared to $60 \pm 1$ nm with STED microscopy (Figure 2D, Supplementary Table 1; Quantification from 110 SVs; Wilcoxon matched-pairs signed rank test, $p<0.0001$;). Similarly, the diameter of VAChT immunofluorescent spots significantly decreased from 173 ± 2 nm in confocal microscopy to 67 ± 1 nm with STED microscopy (Supplementary Figure 1; Supplementary Table 1; Quantification from 110 SVs; Wilcoxon matched-pairs signed rank test, $p<0.0001$).

The first step for analyzing the distribution of VAChT and VGLUT3 immunofluorescent spots from STED images included deconvolution. We checked that the deconvolution of STED images did not introduce artifacts (See Material and methods; Supplementary Figure 2; Supplementary Table 2). As expected, deconvolution increased the resolution of STED images without introducing major alteration of the fluorescent signal as previously reported (Wegel et al., 2016).

The next validation step included control for specificity of immunofluorescent labeling. When SVs were prepared from striatum of VGLUT3$^{-/-}$ mice lacking VGLUT3 (Figure 2F) or when the anti-VGLUT3 antiserum was omitted (Figure 2G), the number of VGLUT3-positive spots was dramatically reduced (Figure 2H, Kruskal Wallis test, Dunn's post hoc test: $p<0.0001$; Supplementary Table 3). From the quantification of VGLUT3 spots in VGLUT3$^{-/-}$ mice, we estimated false positive to represent 5% of the total number of VGLUT3-positive spots. In these three conditions, the number of VAChT-immunopositive spots remained constant (Figure 2H, Supplementary Table 3). The specificity of VAChT labeling was assessed by the disappearance of the signal in VAChT null mice (Janickova et al., 2017). These data confirm the specificity of immunolabeling.

Manual quantification of STED images showed that only 6% of VAChT and VGLUT3 immunopositive spots were strictly superposed (Figure 2I). This observation suggests that, in purified vesicles, a minority of VGLUT3 and VAChT molecules are located at distance smaller than the resolution of STED microscopy (50 nm; Hell and Wichmann, 1994).

To better estimate the distribution of the distance between VGLUT3 and VAChT in striatal purified SVs (Figure 3A-D), we quantified NNDs between the center of the fluorescent spots for VAChT and their closest VGLUT3-fluorescent spot. Interestingly, the frequency histogram showed a first peak in a NND range between 0 to 120 nm with a maximum at 40 ± 5 nm (Figure 3A). In addition to this peak, we also noticed a more spread-out low frequency distribution above 120 nm of NNDs (Figure 3A). Therefore, if VGLUT3 and VAChT are present on the same SV, it can be estimated that the distance between immuno-labeled transporters could be constrained between 0 and 95 nm (Figure 1H,3A-C, and see above for the choice of 95 nm as a cut-off). Quantification showed that 43% of VAChT-positive spots displayed a NND below 95 nm with their closest VGLUT3-immunopositive spot (Figure 3B,C,E). These data suggest that 43% of VAChT and VGLUT3 molecules could be constrained within a distance compatible with a localization of both transporters on the same SV.

We then wondered whether the large pic (0-120 nm) observed with purified SVs (Figure 3A) could be due or not to cluterization (Huttner et al., 1983). CIN varicosities have a mean diameter of 500 nM and contain and average of 10 VAChT-positive spots and 8 VGLUT3-positive spots (Figure 1C,D'). If SVs originate from the same varicosity, all VAChT and VGLUT3 spots should present NNDs smaller than 500 nm (clusterization model Supplementary Figure 3A). Alternatively, if SVs in our preparations are independent or dispersed, in agreement with our EM observations (Figure 2A), VAChT-VGLUT3 NNDs are larger than 500 nm (dispersion model Supplementary Figure 3B). In order to validate the clusterized or the dispersed distribution of VAChT- or VGLUT3-immunospots we tested the clusterization model and the dispersion model (Supplementary Figure 3). We calculated, on isolated SVs preparations, NNDs for each VAChT-immunofluorescent spot with its first, second, third and fourth closest VGLUT3-immunofluorescent spots in a radius of 500 nm, using the DiAna plugin (Gilles et al., 2017). The clusterization model predicted that close to 100% of VAChT spots have a NND < 500 nm with the first to fourth closest VGLUT3 spot (Supplementary Figure 3A). With the dispersion model, decreasing proportions for NNDs < 500 nm from the first to fourth closest VGLUT3 spot would be expected (Supplementary Figure 3B). As shown in Supplementary Figure 3C, relatively to VAChT, 64% of the first closest VGLUT3 spots were located in a radius of 500 nm. The proportion of second, third and fourth closest VGLUT3 spots are only 21%, 7% or 3%, respectively (Quantification from 1506 pairs of VAChT-VGLUT3 spots, from 3 experiments). These data are thus in favor of the dispersion model. Therefore, in our SVs preparation, a majority of immunofluorescent spots correspond to transporters located in independent varicosities. This result is well in line with EM imaging of our SVs preparations.

In the striatum, VAChT and VGLUT3 are present in the same CINs axonal varicosities. In contrast, VGLUT1 or VGLUT2 are expressed in two independent sets of glutamatergic terminals and cannot be expressed by VGLUT3-positive terminals and thus by VGLUT3-positve SVs (Herzog et al., 2001). We therefore estimated NNDs between VGLUT1 and VGLUT3 and between VGLUT2 and VGLUT3. Relatively to VGLUT3-positive spots, both VGLUT1-positive and VGLUT2-positive fluorescent spots appeared evenly distributed in our striatal SVs preparations (Figure 3A,F,H). We observed that 13% and 5% of VGLUT1- or VGLUT2-fluorescent spots were less than 95 nm away from their closest VGLUT3-fluorescent spot (Figure 3G,I), respectively. Statistical analysis showed that the frequency distribution of the NNDs between VAChT and VGLUT3 immunofluorescent spots was significantly different from the one observed between VGLUT1 and VGLUT3 or between VGLUT2 and VGLUT3 immunofluorescent spots (Figure 3A; Kolmogorov-Smirnov test: $p<0.0001$; Supplementary Table 4).

Altogether, these data suggest that a subset of cholinergic SVs co-express both VAChT and VGLUT3 whereas other SVs express only VAChT or only VGLUT3 (Figure 3J). Based on quantification from Figure 2H and 3A,E, we estimate that VAChT and VGLUT3 are both present in 34% of CINs SVs whereas VGLUT3 alone and VAChT alone are found in 26% and 40% of CINs SVs, respectively (Figure 3J; see Supplementary Method 1).

**Discussion**

Several reports suggest that subpopulations of neurons have the ability to release multiple classic neurotransmitters (Varga et al., 2009; El Mestikawy et al., 2011; Shabel et al., 2014). CINs are among this subpopulation of "bilingual" neurons since they express VAChT and VGLUT3 and have demonstrated the capacity to release both ACh and Glut (Higley et al., 2011; Nelson et al., 2014). However, whether ACh and Glut are packaged in the same or in separate pools of SVs and hence are released simultaneously and/or independently remains an open question (for review El Mestikawy et al., 2011; Trudeau and El Mestikawy, 2018).

In the present study, we used super resolution STED microscopy and NND-based quantification to evaluate the relative distribution of VAChT and VGLUT3 in SVs from CINs.

**CINs axonal varicosities contain three subpopulations of SVs expressing VAChT, VGLUT3 or both.**

The limited spatial resolution (200 nm) of conventional fluorescent microscopic approaches is not compatible with precise imaging of proteins in organelles as small as SVs (50 nm). This limitation can be overcome by fluorescence super-resolution microscopies like STED that allow the detection of multiple proteins at a resolution compatible with the size of SVs (Hell and Wichmann, 1994; Choquet et al., 2021; Upmanyu et al., 2022). In this study, we used

STED super-resolution microscopy to perform a detailed analysis of the relative distribution of VAChT and VGLUT3.

We observed that in striatal sections, VAChT and VGLUT3-immunofluorescent spots observed under STED microscopy, rarely overlap. Using the quantification of NNDs between VAChT and VGLUT3, we observed a broad peak of distribution of the NNDs between VAChT and VGLUT3 centered around 40 nm. Since the broad peak spans from 0 to ≈300 nm (Figure 1G), it is difficult to reliably estimate the size of this subpopulation of VAChT/VGLUT3-immunopositive SVs. This difficulty can be due to the fact that, inside CINs varicosities, SVs are tightly packed in SV clusters that may interfere with the analysis of the distances separating VAChT and VGLUT3. To overcome this difficulty, we used isolated striatal SVs and checked with SEM and STED (clusterization vs dispersion model analysis), that our SVs were well separated.

Using the NND-based quantification between VAChT and VGLUT3 immunofluorescent spots, we observed that among all VAChT spots, 43% of them are present located at 0 - 95 nm with a maximal frequency value at 46 nm. This proximity between VAChT and VGLUT3 could be due either to transporter located on the same SV or on independent SVs. To examine this specific point, we applied NND analysis to VGLUT1/VGLUT3 and VGLUT2/VGLUT3 distributions; VGLUT1 and VGLUT2 being expressed in distinct axon terminals and SVs than VAChT and VGLUT3. The frequency distributions of the NNDs between VGLUT1 and VGLUT3 or between VGLUT2 and VGLUT3 is more evenly spread than the one between VAChT and VGLUT3. However, some pairs of spots for vesicular transporters were less than 95 nm away (13% for VGLUT1/VGLUT3 and 5% for VGLUT2/VGLUT3). Therefore, a small part of VAChT-spots and VGLUT3-spots within the 0 - 95 nm peak (5-13% of SVs) may correspond to transporters located on overlapping or very close SVs rather than on the same SV. Taking into account this "background labelling", we estimate that ≈34% of CINS SVs express both VAChT and VGLUT3, ≈26% VGLUT3 alone and ≈40% VAChT alone (see Supplementary Method 1, Figure 3J). Therefore, in theory, 40% of CINs SVs have the capacity to accumulate and release only ACh, more than a quarter only Glut and one third simultaneously ACh and Glut. This distribution is well in line with findings that cholinergic varicosities from the IPN differentially release ACh or Glut depending on their firing activity (Ren et al., 2011). Together, our findings reveal an unsuspected level of sophistication in the sorting of VAChT and VGLUT3 in CINs varicosities. This is interesting since CINs have different patterns of firing when they signal for reward (Atallah et al., 2014). Based on findings in the IPN, it can be proposed that CINs could preferentially release Glut during tonic activity (low frequency) and ACh and/or Glut during burst activity (high frequency). Importantly, ACh and Glut released by CINs have opposite effects on dopamine effluxes. Given the crucial role of dopamine to shape reward-guided behaviors (Morris et al., 2004; Cragg, 2006; Frahm et al., 2015), an important challenge for the coming years will be to identify when these different modes of cholinergic and glutamatergic transmission are used by CINs.

Previous reports show that the presence of VGLUT3 increases the vesicular accumulation of ACh in CINs through vesicular synergy (Gras et al., 2008; El Mestikawy et al., 2011). The current mechanistic explanation of vesicular synergy is based on an increased acidification of cholinergic SVs due to concomitant VGLUT3-dependent Glut uptake. (Gras et al., 2008; El Mestikawy et al., 2011). Therefore, vesicular synergy implies that VAChT and VGLUT3 are expressed on the same population of SVs. Our present data bring further support and morphological basis for the current mechanistic understanding of vesicular synergy. Interestingly, we previously reported that VGLUT3, through vesicular synergy, was able to increase ACh vesicular loading by a factor of two-three folds (Gras et al., 2008). Our data showing that only a third of SVs co-express VAChT and VGLUT3, suggest that vesicular synergy within this subpopulation of SVs could probably be up to a tenfold increase. The existence and functions of such a subpopulation of VAChT/VGLUT3-positive SVs putatively containing very high amounts of ACh remains to be established.

Interestingly, a recent study using SVs from the whole brain demonstrated that only 1.5% of SVs express both VAChT and VGLUT3 (Upmanyu et al., 2022). Our data suggest that striatal SVs co-expressing VAChT and VGLUT3 are 20 times more abundant than in total brain extracts. It is well established that most of the brain ACh is found in the striatum. Furthermore, in this structure CINs form a dense network of varicosities and cholinergic fibers represent ≈ 15% of striatal terminal/varicosities. It is therefore not surprising that VAChT/VGLUT3-positive SVs are more abundant in the striatum than in total brain extracts. Interestingly, total brain SVs show a high level of colocalization of VGLUT1 with ZnT3 (the $Zn^{2+}$ transporter, (Upmanyu et al., 2022)) and $ZN^{2+}$ positive ions facilitate the vesicular accumulation of negatively charged glutamate (Upmanyu et al., 2022). This observation confirms and extends the concept that multiple transporters and accumulation of "counter ions" could be indispensable for vesicular synergy to occur (El Mestikawy et al., 2011). However, it should be mentioned that VGLUT-operated vesicular synergy has been proposed to increase ACh or DA vesicular concentration by augmenting the ΔpH (Gras et al., 2008; Hnasko et al., 2010), whereas ZnT3 and $Zn^{2+}$ vesicular accumulation should accelerate glutamate uptake by increasing ΔΨ.

**Conclusion**

Our data suggest the existence of three subpopulations of SVs expressing VAChT and/or VGLUT3. This unexpected heterogeneity of SVs in CINs varicosities may confer complex releasing properties to CINs at various firing frequencies and may thus provide CINs with a sophisticated level of regulation of striatal functions. These findings open new avenues to better understand striatal functions and associated pathologies.

**Conflict of interest statement**

The authors declare that the research was conducted in the absence of any commercial or financial relationships that could be construed as a potential conflict of interest.

**Author Contributions**

N.P. and S.E.M and V.B. participated in the research design; P.C., M.D., O.P., A.H., SD, N.P. and V.B. conducted experiments; S.D. performed FISH experiments; V.B. and L.D. performed nearest neighbor distance analysis of STED fluorescent spots; P.C., M.D., L.D., N.P., S.E.M. and V.B. performed data analysis; V.B., M.D., E.H., S.B., D.G. and J.F.G. contributed to imaging and data analyses; M.T., A.C. and A.G. contributed to electron microscopy experiments; V.B., N.P., E.H., V.F., S.D. and S.E.M. contributed to the writing or proofreading of the manuscript; all authors red and approved the final version of the manuscript.

**Fundings**

This research was financially supported by INSERM, CNRS, *Sorbonne Université*, SATT-Lutech (Paris), the "Agence Nationale pour la Recherche" (ANR,

ANR- 13- SAMA- 0005- 01), Natural Sciences and Engineering Research Council Discovery Grants (RGPIN/386431-2012 and RGPIN/04682-2017) and by UNAFAM, the Bouygues group with the scientific expertise provided by the FRC's scientific council and 80PRIME CNRS and FRC grant. The Imaging facility of the *Institut de Biologie Paris- Seine* was supported by the Ile-de-France region, CNRS, GIS-IBISA and Sorbonne University.

**Acknowledgments**

We thank France Lam from the Imaging facility of the *Institut de Biologie Paris- Seine* for confocal imaging assistance. STED imaging was carried out at NeurImag Imaging core facility, part of the IPNP, Inserm 1266 unit and Université de Paris. We thank Leducq foundation for funding the Leica SP8 Confocal/STED 3DX system and NeurImag imaging core facility for their scientific expertise in data acquisition, processing, and analysis. We thank Philippe Bun from NeurImag Imaging core facility (Université de Paris, IPNP, INSERM U1266, France) for assistance with STED microscopy. We thank Marie-Laure Niepon at the Image platform at Institute de la Vision (Paris, France) for slide scanning.

**Conflict of interest statement**

The authors declare that the research was conducted in the absence of any commercial or financial relationships that could be construed as a potential conflict of interest.

**Data Availability**

Any raw data supporting the current study is available from the Lead Contact (veronique.bernard@inserm.fr) upon request and will be made available upon reasonable request.

**Figures legends**

**Figure 1: Immunofluorescent detection of VAChT and VGLUT3 in the mouse dorsolateral striatum (DLS) with confocal and STED microscopy.** **(A,B)** VAChT and VGLUT3 are co-detected in the cell body of a striatal neuron *(\* in* **A***)* and in axonal varicosities (arrows in **B**). **(C,C')** The same axonal varicosity observed by confocal microscopy **(C)** and by STED microscopy **(C')**. **(D,D')** Axon and varicosities (arrows in **D**) of a CINs observed with STED microscopy. **(E)** Quantification of VAChT-, VGLUT3- and VAChT/VGLUT3-immunopositive spots per varicosity surface. **(F)** Manual quantification of the distribution per varicosity of VAChT, VGLUT3 and VAChT/VGLUT3 immunopositive spots. **(E,F)** Quantification from 7 mice, 3 sections per animal, 155 varicosities, 2883 spots in total. **(G)** Batch analysis of NND between VAChT- and VGLUT3-immunofluorescent spots by STED microscopy per 10 nm bins. **(H)** Schematic representation to estimate the cut-off for considering that VAChT and VGLUT3 are present or not on the same SV. This estimation is based on the steric hindrance of the primary/secondary antibody complex (45 nm) and the mean diameter of the SV (50 nm). **(I)** Percentage of VAChT spots displaying a NND with their closest VGLUT3 spot above and below 95 nm. **(G,I)** Quantification from 5 microscopic fields (625 μm$^2$), analysis of 914 pairs of immunofluorescent spots for VAChT and VGLUT3. VAChT-VGLUT3 pairs were identified in a radius of 1000 nm. Data are expressed as means ± SEM. **(A,D')** Fluorochromes : Alexa 594 (VAChT, green) and Abberior Star 635P (VGLUT3, purple).

**Figure 2: Immunohistochemical detection of VAChT and VGLUT3 on isolated mouse striatal synaptic vesicles with confocal, STED and electron microscopy.** **(A)** Scanning electron microscopy (SEM) of striatal SVs covered with a 5nm layer of platinium. **(B)** Quantification of the mean diameter of SVs with on SEM images (n=484 SVs). **(C,C')** The

same isolated SVs immunofluorescent for VGLUT3 observed by confocal microscopy **(C)** and by STED microscopy **(C')**. One spot observed under the confocal microscope correspond to three spots when observed under the STED microscope (inset in **C,C'**). **(D)** Quantification of the mean diameter of VGLUT3 fluorescent spots with confocal or STED microscopy (n=110 from 20 animals). Wilcoxon matched-pairs signed rank test, **p*<0.0001. **(E-G)** Simultaneous immunofluorescent detection with STED microscopy of VAChT- and VGLUT3-immunofluorescent spots on isolated mouse striatal SVs from VGLUT3$^{+/+}$ mice **(E,G)** or VGLUT3$^{-/-}$ mice **(F)** in the presence **(E,F)** or absence **(G)** of anti-VGLUT3 antiserum. **(E)** In VGLUT3$^{+/+}$ vesicles, some VAChT- and VGLUT3-immunofluorescent spots appear isolated (arrows in **E**) whereas other spots are close (arrowheads and inset). **(H)** Automatized quantification of the number of VAChT-, VGLUT3- and VAChT+VGLUT3-immunofluorescent spots by surface of coverslip from n=19 microscopic fields (625 μm$^2$) from vesicle preparation from 20 animals in three independent experiments. Kruskal-Wallis test followed by the Dunn's post-hoc test, ** *p* <0.0001. **(I)** Manual quantification of VAChT-, VGLUT3 or VAChT/VGLUT3-immunopositive spots per surface of coverslip. Data are expressed as means ± SEM. **(C,C',E-G)** fluorochromes : Alexa 594 (VAChT) and Abberior Star 635P (VGLUT3).

**Figure 3: Quantitative analysis of the NNDs between VAChT, VGLUT1 or VGLUT2 and VGLUT3-immunofluorescent spots on striatal isolated SVs.**

**(A)** Quantification and frequency histograms of the NND between the centers of VAChT, VGLUT1 or VGLUT2-immunofluorescent spots and their closest VGLUT3 spots. A total of 7048, 6017 and 6827 pairs of VAChT/VGLUT3, VGLUT1/VGLUT3 and VGLUT2/VGLUT3 immunofluorescent spots, respectively, were analyzed in 3 independent experiments. **(B-D')** Examples of VGLUT3- and VAChT-fluorescent spots superposed **(B)**, juxtaposed **(C)** or separated **(D). (F,H)** Examples of VGLUT1/VGLUT3 **(F)** and VGLUT2/VGLUT3 **(H)** fluorescent spots. **(E,G,I)** Percentage of VAChT, VGLUT1 or VGLUT2 spots displaying a NND with their closest VGLUT3 spot above and below 95nm. **(J)** Hypothetical model of the distribution of VGLUT3 and VAChT in SVs from CINs varicosities and estimation of the proportion of VAChT+VGLUT3-, VAChT- alone and VGLUT3- alone SVs in CINs varicosities. **(B,D,F,H)** fluorochromes : Alexa 594 (VAChT,VGLUT1,VGLUT2) and Abberior Star 635P (VGLUT3).